\documentstyle[prl,aps,twocolumn,floats,graphicx,amsfonts,amssymb,%
amsmath]{revtex}
\begin{document}
\draft
\preprint{}
\title{Collective Excitations of Dilute Bose-Fermi Superfluid Mixtures}
\author{Bernhard Hikaru Valtan\thanks{e-mail:
valtan@kh.phys.waseda.ac.jp},
 Munehiro Nishida and Susumu Kurihara}
\address{
Department of Physics, Waseda University,\={O}kubo, Shinjuku,
Tokyo 169-8555, Japan}
\date{\today}
\twocolumn[\hsize\textwidth\columnwidth\hsize\csname@twocolumnfalse%
\endcsname        
\maketitle
\begin{abstract}
  Using the effective action formalism,
we investigate collective excitations of a dilute mixture of a 
Bose gas and a two component Fermi gas when 
both bosons and fermions have undergone superfluid transitions. We show
that there is a repulsion between Bogoliubov and Anderson modes,
which has important implications including disappearance
of boson superfluidity.
We derive an analytic expression for the  long-wavelength
dispersion relation of the mixture at zero temperature 
and give a condition for the instability.
We also numerically calculate the damping rate due 
to boson-fermion interaction at finite temperatures
and show that the two modes are stable  
at sufficiently low temperatures. 
\end{abstract}
\pacs{%67.40.Hf, 64.60.Qb, 67.40.Fd
} 
]
{
 \renewcommand{\thefootnote}%
      {\fnsymbol{footnote}}
    \footnotetext[1]{e-mail: valtan@kh.phys.waseda.ac.jp}
  }
\narrowtext
\section{Introduction}
Experimental realization of Bose-Einstein condensation (BEC)
 \cite{BEC} of trapped atomic Bose gases 
and achievement of trapped quantum 
degenerate Fermi gas (DFG) \cite{Jin99} 
are two of the most important  prospers 
in the field of ultracold atomic gas.
Recently, $^{40}\mathrm{K}$ atoms has been cooled 
to one-fifth of the Fermi temperature $(0.2 T_F)$ \cite{Jin01}.
One of the cardinal goals for the cooling of 
Fermi gas is the observation of BCS-type superfluidity 
 \cite{StoofLi6}. 
Except for $^6\mathrm{Li}$ which has an enormously
large negative scattering length \cite{2160}, creation 
of effective attractive interaction for pairing is necessary,
methods such as Feshbach resonance \cite{feshLi,feshRb,feshK}
and phonon exchange in Fermi-Bose mixtures have been proposed
\cite{StoofBF}. 

Sympathetic  cooling of atomic Bose-Fermi gas mixtures has also risen
to a thriving field. Since Bose gas can act as a coolant
for fermions, severe decrease of rethermalization 
due to Pauli blocking 
can be avoided. In addition, this leads us to
a fascinating system of dilute Bose-Fermi mixture to explore. 
Success in simultaneous trapping of Bose-Fermi isotopic 
mixture has been reported \cite{twoLitrap,twoRbtrap} and 
simultaneous quantum degeneracy at temperature as low as
$ 0.25 T_F$ has been observed in bosonic $^7\mathrm{Li}$ and 
fermionic $^6\mathrm{Li}$ mixture \cite{Hulet}. Density profiles 
in harmonic traps \cite{Molmer,Tosi} and collective excitations 
\cite{StoofBF,Pu,Capuzzi} of Bose-Fermi mixtures has been investigated
so far. 

Collective excitations of superfluid Fermi gas are predicted
to be detectable \cite{Baranov,Minguzzi} and sound propagation
in BEC has already been observed \cite{BECSound}. It would 
be interesting to investigate how these two sound modes 
act upon each other.
In this paper, we study the collective excitations of dilute 
Bose-Fermi mixture when both boson and fermion have undergone 
superfluid transition. Spatial homogeneity is assumed throughout 
the calculation since overlapping of bosons and fermions can
only be large in a gradual or box-like trap. 
Based on the effective action formalism,
long-wavelength  dispersion relation of the superfluid 
mixture  at zero temperature is derived  and damping rates due to 
coupling of the Bogoliubov and Anderson mode are also obtained
by analytic continuation of the RPA-polarization bubbles. 
Instability of the superfluid modes in this mixture  
is also discussed.
\section{Formulation}
\subsection{Model}
We shall start from a imaginary time path integral 
for which the grand canonical partition function 
reads
\begin{eqnarray}
\mathcal{Z} &=&
 \int {\mathcal D} \phi^\ast {\mathcal D } \phi {\mathcal D } \psi^\ast {\mathcal D }\psi
\exp \bigg\{ -\frac{1}{\hbar} (S_B[\phi^\ast,\phi ]
  \nonumber\\  &&
  + S_F[\psi^\ast,\psi] +  S_I[\phi^\ast,\phi,\psi^\ast,\psi]  ) 
       \bigg\} ,
\end{eqnarray}
where the total action consists of terms representing the Bose gas,
\begin{eqnarray}
S_B[\phi^\ast,\phi] &=& \int_0^{\hbar\beta} d\tau \int d {\boldsymbol x}  \nonumber\\ &&
 \times  \bigg\{ \phi^\ast({\boldsymbol x},\tau) \bigg(\hbar\frac{\partial}{\partial \tau} -
\frac{\hbar^2 \nabla^2}{2 m_B} - \mu_B \bigg) \phi({\boldsymbol x},\tau)
\nonumber\\ &&
 +\frac{g_B}{2}| \phi({\boldsymbol x},\tau) |^4 \bigg\} ,
\end{eqnarray}
the Fermi gas,
\begin{eqnarray}
S_F[\psi^\ast,\psi]&=& \sum_{\alpha} \int_0^{\hbar\beta} d\tau \int d {\boldsymbol  x} 
 \nonumber\\ &&
\times \bigg\{ \psi^\ast_\alpha({\boldsymbol  x },\tau) \bigg(\hbar\frac{\partial}{\partial \tau} -
\frac{\hbar^2 \nabla^2}{2 m_F} - \mu_\alpha \bigg) \psi_\alpha({\boldsymbol x },\tau)
\nonumber\\ &&
 + \frac{g_F}{2} |\psi_\alpha({\boldsymbol x },\tau) |^2  | \psi_{- \alpha } ({ \boldsymbol x },\tau) |^2 
 \bigg\} ,
\end{eqnarray}
and  boson-fermion interaction
\begin{eqnarray}
&& S_I[\phi^\ast,\phi, \psi^\ast,\psi]   \nonumber\\
&&= \sum_{\alpha} \int_0^{\hbar\beta} d\tau \int d {\boldsymbol  x}
~ g_\alpha |\psi_\alpha({ \boldsymbol  x},\tau) |^2  |\phi({ \boldsymbol x},\tau) |^2 ,
\end{eqnarray}
where $\phi(\mathbf{x},\tau)$ are complex fields 
and  $\psi_\alpha(\mathbf{x},\tau)$ are Grassman fields 
describing  Bose component and the two hyperfine Fermi 
component $\alpha =  \{ \uparrow, \downarrow \} $ 
respectively. $g_x =\frac{2 \pi \hbar^2 a_x}{ m_x}$ is the
coupling constant, for which $a_B$, $a_F$ and
 $a_{\uparrow, \downarrow}$ are the scattering length 
of boson-boson, fermion-fermion and boson-fermion interaction respectively. 
Note that only interaction between two different hyperfine 
states is considered since Pauli principle prohibits
s-wave scattering between two identical fermions. 
$m_x =\frac{m_1 m_2}{m_1 + m_2} $ is the 
reduced mass and $\mu_x$ is the chemical potential 
of the corresponding component.

  In order to introduce an auxiliary field 
corresponding to BCS pairing field, we perform
a Stratonovich-Hubbard transformation to the 
fermion-fermion interaction,  
\begin{eqnarray}
&&
\exp \bigg\{-\frac{g_F}{\hbar} \int_0^{\hbar\beta} d\tau \int d {\boldsymbol x}
 |\psi_\uparrow({\boldsymbol x},\tau) |^2  | \psi_{\downarrow } ({\boldsymbol x},\tau) |^2 \bigg\}
 \nonumber \\ 
 &=& \int{\mathcal D}\Delta^\ast { \mathcal D} \Delta\exp \bigg\{ \frac{1}{\hbar}
 \int_0^{\hbar\beta} d\tau \int d { \boldsymbol x}
  \bigg[ \frac{|\Delta({\boldsymbol x},\tau) |^2}{g_F }  \nonumber\\   &&
+ \Delta^\ast ({\boldsymbol x},\tau) \psi_\downarrow({\boldsymbol x},\tau) \psi_\uparrow({\boldsymbol x },\tau)
 \nonumber \\ &&
+  \psi_\uparrow^\ast({\boldsymbol x},\tau) \psi_\downarrow^\ast({\boldsymbol x},\tau) 
\Delta ({\boldsymbol x},\tau)
 \bigg] \bigg\}.
\end{eqnarray}
Integrating over the fermionic  field, we obtain
an effective action,  
\begin{eqnarray}
S_{\mathit eff }[\Delta^\ast,\Delta,\phi^\ast,\phi] = \int_0^{\hbar\beta} d\tau \int d {\boldsymbol x}
 \bigg[ - \frac{|\Delta({\boldsymbol x },\tau) |^2}{g_F} \bigg]   \nonumber\\
+ \  S_B[\phi^\ast,\phi] - \hbar {\mathrm Tr}[ { \mathrm ln}(-{ \mathcal G}^{-1})] ,
\end{eqnarray}
for which the inverse of Green's function in the third term 
can be expanded perturbatively, 
\begin{eqnarray}
&&  -{\mathcal G }^{-1}({\boldsymbol x, \tau ;\boldsymbol x^\prime , \tau^\prime  }) =  \nonumber\\
&& \frac{1}{\hbar} 
\begin{pmatrix}
  \hbar \frac{\partial}{\partial \tau} - \frac{\hbar^2 \nabla}{2 m_F}
- \mu_{\downarrow} + g_{\downarrow} |\phi |^2
& \Delta \\
\Delta^\ast & 
\hbar \frac{\partial}{\partial \tau} + \frac{\hbar^2 \nabla}{2 m_F}
+ \mu_{\uparrow} - g_{\uparrow} |\phi |^2
\end{pmatrix}  
\nonumber \\
&& \times \delta({\boldsymbol x - \boldsymbol x ^\prime} )\delta(\tau - \tau^\prime ).
\end{eqnarray}
\subsection{Perturbation Expansion}
Since the trace of (7) is invariant under unitary transformation,
we can perform a rotation in Nambu space,
$U(\theta) = e^{-\frac{i}{2}\sigma_3\theta({\boldsymbol x},\tau)}$,
which corresponds to a gauge transformation.
Separating the pairing field into phase and amplitude, i.e.
$\Delta({\boldsymbol x},\tau) =\Delta_a({\boldsymbol x},\tau)
 e^{i \theta_p({\boldsymbol x},\tau) }$, and choosing $\theta =
 \theta_p$ leads to a new inverse of Green's function  
$\tilde{\mathcal G}^{-1}$ with a real energy gap. 

Bogoliubov approximation is then applied to the Bose field 
and the pairing field is also separated into its average 
and fluctuation. i.e. 
\begin{eqnarray}
\phi({\boldsymbol x},\tau) = \sqrt{n_B} + \phi^\prime({\boldsymbol x},\tau), \\
\Delta_a({\boldsymbol x },\tau) = \Delta_0 + \delta({\boldsymbol x},\tau).
\end{eqnarray}

Assuming quantum fluctuation to be small, the third term in (6) can be
expanded,
\begin{eqnarray}
- \hbar {\mathrm Tr}[ { \mathrm ln}(-{\tilde{\mathcal G}}^{-1})] =
 - \hbar  { \mathrm Tr}[{\mathrm ln}(-{ \tilde{\mathcal G}}_0^{-1})]
+  \hbar\sum_{j=1}^\infty \frac{1}{j} {\mathrm Tr}[({ \tilde{\mathcal G}}_0\Sigma)^j],
\end{eqnarray}
where the fluctuation part reads,
\begin{eqnarray}
\Sigma({\boldsymbol x},\tau) &=& - \frac{1}{\hbar} 
\begin{pmatrix}
K + L + M
& \delta({\boldsymbol x},\tau) \\
\delta({\boldsymbol x},\tau)   & 
-K + L + M 
\end{pmatrix}
 \nonumber\\
&=& - \frac{1}{\hbar} \left(
K \sigma_3 + ( L + M ) \sigma_0 + \delta({\boldsymbol x},\tau)  \sigma_1
\right),
\end{eqnarray}
with,
\begin{eqnarray}
K &=& i \gamma( {\boldsymbol  x},\tau) + \frac{m_F v^2( {\boldsymbol  x},\tau)}{2} 
   + g_A\left[\sqrt{n_B}(\phi^{\prime}( {\boldsymbol  x},\tau) + \phi^{\prime\ast}( {\boldsymbol  x},\tau) ) 
   \right.     \nonumber \\
&& \left.  + \phi^{\prime}( {\boldsymbol  x},\tau) \phi^{\prime\ast}( {\boldsymbol  x},\tau) )  \right],  \\
L &=& \frac{i \hbar}{2} [ \nabla \cdot   {\boldsymbol v}( {\boldsymbol  x},\tau) 
+ {\boldsymbol v}( {\boldsymbol  x},\tau) \cdot  \nabla
], \\
M &=& g_D\left[\sqrt{n_B}(\phi^{\prime}( {\boldsymbol  x},\tau)+ \phi^{\prime\ast}( {\boldsymbol  x},\tau) ) 
+ \phi^{\prime}( {\boldsymbol  x},\tau) \phi^{\prime\ast}( {\boldsymbol  x},\tau) )  \right], 
\end{eqnarray}
in which 
${\boldsymbol v}({\boldsymbol x},\tau) = \frac{\hbar}{2 m_F} \nabla
\theta({\boldsymbol x},\tau)$ and 
$\gamma({\boldsymbol x},\tau) = \frac{\hbar}{2}\dot{\theta}({\boldsymbol x},\tau)$
are the superfluid velocity and chemical potential for Cooper pairs respectively.
The average of boson-fermion interaction,
$g_A = \frac{g_\uparrow+g_\downarrow}{2}$
and the deviation of interaction strength
from average $g_D = \frac{g_\uparrow-g_\downarrow}{2}$
has also been introduced 
in order to rewrite (11) in Pauli matrices.

Since we are dealing with a homogeneous system,
it is convenient to perform a Fourier transformation.
The unperturbed Green's function reads,
\begin{eqnarray}
{\tilde{\mathcal G}}_0({ \boldsymbol k},\omega_n) &=& \frac{\hbar}{D({ \boldsymbol k},\omega_n)} 
\begin{pmatrix}
i \hbar \omega_n + \epsilon_\uparrow({ \boldsymbol k})
& \Delta_0 \\
\Delta_0 & 
i \hbar \omega_n - \epsilon_\downarrow({ \boldsymbol k})
\end{pmatrix} 
 \nonumber\\
&\equiv& \begin{pmatrix}
{\mathcal G}_\downarrow({ \boldsymbol k},\omega_n)
& {\mathcal F}({ \boldsymbol k},\omega_n)  \\
{ \mathcal F}({ \boldsymbol k},\omega_n)  & 
{ \mathcal G}_\uparrow({ \boldsymbol k},\omega_n)
\end{pmatrix}
,
\end{eqnarray}
with,
\begin{eqnarray}
D({ \boldsymbol k},\omega_n) = (i \hbar \omega_n -\epsilon_\downarrow({ \boldsymbol k}))
(i \hbar \omega_n + \epsilon_\uparrow({ \boldsymbol k})) - |\Delta_0|^2 ,
\end{eqnarray}
where 
$\epsilon_\alpha({\boldsymbol k}) = \frac{\hbar^2 {\boldsymbol k}^2}{2m} - \mu_\alpha + g_\alpha n_B$
and $ \omega_n = \frac{2 \pi n}{\hbar \beta }$ is the even Matsubara
frequency.
 
Expanding (10) to the second order of fluctuation around its minimum
leads to the dispersion relation of collective modes.  
The requirements of the linear terms of fluctuations to be zero
$( \delta S^{(1)} = 0 )$  are the well known Hugenholz-Pines relation 
\begin{eqnarray}
\mu_B =  g_B n_B +g_\downarrow n_\downarrow +g_\uparrow n_\uparrow ,
\end{eqnarray}
with $  {\mathcal G }_{\alpha}({\boldsymbol x, \tau ;\boldsymbol x , \tau  }) =
n_\alpha  $ for the Bose field and BCS gap equation
\begin{eqnarray}
{\mathcal F}({\boldsymbol x,\tau; \boldsymbol x,\tau}) = \frac{\Delta_0}{g_F} , 
\end{eqnarray}
for the pairing field. 
\subsection{RPA Polarization Bubbles}
Expanding (10) to the second order gives rise to 
various kinds of RPA Bubbles :
\begin{eqnarray}
f_0({\boldsymbol k},\omega_n) &=& 
\frac{1}{\hbar \beta V} \sum_{{\boldsymbol p},m} {\mathcal F}({\boldsymbol k},\omega_n)
{\mathcal F}({\boldsymbol p +\boldsymbol k},\omega_m+\omega_n) , \\
g_0({\boldsymbol k},\omega_n) &=& 
\frac{1}{\hbar \beta V} \sum_{{\boldsymbol p}  ,m} {\mathcal G} _\downarrow({\boldsymbol k},\omega_n)
{\mathcal G}_\downarrow({\boldsymbol p +\boldsymbol k},\omega_m+\omega_n) ,  \\
h_0({\boldsymbol k},\omega_n) &=& 
\frac{1}{\hbar \beta V} \sum_{ {\boldsymbol p} ,m} {\mathcal G} _\downarrow({\boldsymbol k},\omega_n)
{\mathcal G}_\uparrow({\boldsymbol p +\boldsymbol k},\omega_m+\omega_n) ,  \\
k_0^+({\boldsymbol k},\omega_n) &=& 
\frac{1}{\hbar \beta V} \sum_{{\boldsymbol p}  ,m} {\mathcal G} _\downarrow({\boldsymbol k},\omega_n)
{\mathcal F}({\boldsymbol p +\boldsymbol k},\omega_m+\omega_n) ,
\end{eqnarray}
note that inverting  ${\mathcal G}$ and ${\mathcal F}$ at (22)
gives   $k_0^-({\boldsymbol k},\omega_n)$.

The remaining bubbles such as :
\begin{eqnarray}
g_1({\boldsymbol k},\omega_n) && \nonumber\\
=  \frac{1}{\hbar \beta V} \sum_{ {\boldsymbol p}   ,m} &\left({\boldsymbol p} + \frac{\boldsymbol  k}{2} \right)&
 {\mathcal G} _\downarrow({\boldsymbol k},\omega_n)
{\mathcal G}_\downarrow({\boldsymbol p +\boldsymbol k},\omega_m+\omega_n) , \\
g_2({\boldsymbol k},\omega_n)  && \nonumber\\
=  \frac{1}{\hbar \beta V} \sum_{ {\boldsymbol p}   ,m} &\left({\boldsymbol p} + \frac{\boldsymbol k}{2} 
\right)^2& {\mathcal G} _\downarrow({\boldsymbol k},\omega_n)
{\mathcal G}_\downarrow({\boldsymbol p +\boldsymbol k},\omega_m+\omega_n) ,
\end{eqnarray}
can be obtained as a combination of (19)-(22) 
by considering small rotation $U(\chi) =
e^{\frac{i}{2}\sigma_3\chi({\boldsymbol x})}$ of
 phase,   since gauge invariance gives rise to a Ward 
identity  \cite{Schrieffer,Otterlo}. 

 The real part of the bubbles can be calculated in
a purely analytical manner \cite{Popov,Otterlo} at absolute zero, 
the result of integration is independent of the 
difference $ g_\uparrow -  g_\downarrow $ and 
$ \mu_\uparrow -  \mu_\downarrow $,
\begin{eqnarray}
&& {\mathrm  Re} f_0 ({\boldsymbol k},\omega) \nonumber\\
&=& \frac{N(0)}{2}\left[
1 - \frac{\hbar^2}{6 \Delta^2_0} \left( - \omega^2 + \frac{v_F^2 k^2}{3}
 \right) + \ldots  \right] , \\
&& {\mathrm  Re} g_0 ({\boldsymbol k},\omega) \nonumber\\
&=& - \frac{N(0)}{2}\left[
1 - \frac{\hbar^2}{6 \Delta^2_0} \left( - \omega^2 - \frac{v_F^2 k^2}{3}
 \right) + \ldots \right] , \\
&& {\mathrm  Re} h_0 ({\boldsymbol k},\omega) \nonumber\\
&=& \frac{1}{g_F} + \frac{N(0)}{2}\left[
1 - \frac{\hbar^2}{3 \Delta^2_0} \left( - \omega^2 + \frac{v_F^2 k^2}{3}
 \right) + \ldots \right] , \\
&& {\mathrm  Re} k_0^\pm ({\boldsymbol k},\omega) \nonumber\\
&=& \pm  \frac{N(0) \hbar \omega}{4\Delta_0}\left[
1 - \frac{\hbar^2}{6 \Delta^2_0} \left( - \omega^2 + \frac{v_F^2 k^2}{3}
 \right) + \ldots \right] .
\end{eqnarray}
  In order to obtain the imaginary part, 
we have to perform an analytic continuation
carefully by representing the Matsubara sum
in the form of a contour integral \cite{AGD}.
The final result is, 
\begin{eqnarray}
{\mathrm Im} &\frac{1}{\hbar \beta V}& \sum_{\boldsymbol p,m}{\mathcal G}_{\downarrow}
({\boldsymbol p},\varepsilon)
{\mathcal F}({\boldsymbol p +\boldsymbol  k},\varepsilon + \omega_n)    \nonumber \\
&=&  \frac{1}{(2 \pi)^4} \int d {\boldsymbol p} \int^\infty_{- \infty} d \varepsilon
{\mathrm Im}G_{\downarrow R} ({\boldsymbol p},\varepsilon)
{\mathrm Im} F_{R}({\boldsymbol p +\boldsymbol  k},\varepsilon + \omega) \nonumber \\ 
&& \times
\left( 
\tanh \frac{\hbar \varepsilon}{2 k_B T} - \tanh \frac{\hbar (\varepsilon +  \omega)}{2 k_B T}
\right) .
\end{eqnarray}

In the limit of long-wavelength, i.e. $\frac{ \hbar \omega}{2 k_B T} \ll 1
$ and
$\frac{ \hbar \omega}{\Delta_0} \ll 1 $, 
the imaginary parts of bubbles can be obtained.
Assuming $ g_\uparrow =  g_\downarrow $ and 
$ \mu_\uparrow =  \mu_\downarrow $ for simplicity, we get 
\begin{eqnarray}
{\mathrm Im}f_0 ({\boldsymbol k},\omega,T) &=&- \frac{ N(0) \pi \Delta^2_0 \omega}{2 k_B T v_F k}
\int^\infty_{\frac{\Delta_0}{\sqrt{1-c^2}}}d E \frac{{\mathrm sech}^2 \frac{E}{2 k_B T}}{E^2-\Delta_0^2} ,
\\
{\mathrm Im}g_0 ({\boldsymbol k},\omega,T) &=& - \frac{ N(0) \pi \omega}{2 k_B T v_F k}
\int^\infty_{\frac{\Delta_0}{\sqrt{1-c^2}}}d E \nonumber \\
&&  \times  \frac{(2 E^2- \Delta_0^2){\mathrm sech}^2 \frac{E}{2 k_B T}}{E^2-\Delta_0^2} ,
\\
{\mathrm Im}h_0 ({\boldsymbol k},\omega,T) &=& {\mathrm Im}f_0 ({\boldsymbol k},\omega,T) ,
\\
{\mathrm Im}f_0 ({\boldsymbol k},\omega,T) &=&
\frac{\Delta_0}{\hbar \omega}
{\mathrm Im}(k_0^+ ({\boldsymbol k},\omega,T) - k_0^- ({\boldsymbol k},\omega,T)) ,
\end{eqnarray}
where $E = \sqrt{\xi^2 +\Delta_0^2}$ and $c = \frac{\omega}{v_F k} < 1$.
(33) can be verified from the Ward identity of $\sigma_0$ rotation \cite{Otterlo}.
The remaining integration is done numerically.
\subsection{Inverse Green's Function of Mixture}
The second order of fluctuation gives us the desired dispersion
relation of the mixture. In analogy to the pairing field, we separate
the Bose field to it's phase and amplitude as well,
\begin{eqnarray}
\phi^\prime ({\boldsymbol k},\omega) =\phi_{\mathrm A}({\boldsymbol k},\omega)
 + i \phi_{\mathrm P}({\boldsymbol k},\omega) .
\end{eqnarray}
Introducing a vector notation,
\begin{eqnarray}
\Psi^\dagger = \Bigl(
\begin{matrix}
\phi_{\mathrm A}({\boldsymbol k},\omega),  &  \phi_{\mathrm P}({\boldsymbol k},\omega), &
 \delta ({\boldsymbol k},\omega), & \theta ({\boldsymbol k},\omega)
\Bigr) ,
\end{matrix}
\end{eqnarray}
the second order of fluctuation can be written in a matrix form,
\begin{eqnarray}
S^{(2)}= \frac{1}{2 \pi V} \sum_{\boldsymbol k} \int d \omega \Psi^\dagger G^{-1} ({\boldsymbol k},\omega,T) \Psi ,
\end{eqnarray}
where $G^{-1}$ is the inverse of Green's function of the mixture,
\begin{eqnarray}
G^{-1} =
\begin{pmatrix}
{\mathrm M}_{\mathrm A A}  &
i\hbar \omega  & {\mathrm M}_{\mathrm A \delta }   & {\mathrm M}_{\mathrm A \theta }  \\
- i \hbar \omega & \frac{\hbar^2 k^2}{2m_B} & 0 & 0  \\
 - {\mathrm M}_{\mathrm A \delta }  & 0 & {\mathrm M}_{\delta \delta } & 0 \\
- {\mathrm M}_{\mathrm A \theta }  & 0 & 0 & {\mathrm M}_{\theta \theta } 
\end{pmatrix}
,
\end{eqnarray}
with matrix elements : 
\begin{eqnarray}
{\mathrm M}_{\mathrm A A} &=& 
\frac{\hbar^2 k^2}{2 m_B} +2 g_B n_B + g_A^2 n_B(g_0({\boldsymbol k})-f_0({\boldsymbol k})) \nonumber \\ 
&&+ g_D^2 n_B(g_0({\boldsymbol k})+f_0({\boldsymbol k})) , \\
{\mathrm M}_{\mathrm A \delta} &=& g_D \sqrt{n_B}(k_0^+({\boldsymbol k})+k_0^-({\boldsymbol k})) , \\
{\mathrm M}_{\mathrm A \theta} &=& -i g_A \sqrt{n_B}(k_0^+({\boldsymbol k})-k_0^-({\boldsymbol k})) , \\
{\mathrm M}_{\mathrm \delta \delta} &=& h_0({\boldsymbol k}) + f_0({\boldsymbol k}) -\frac{1}{g_F}  , \\
{\mathrm M}_{\mathrm \theta \theta} &=&  \Delta_0^2 (h_0({\boldsymbol k}) - f_0({\boldsymbol k}) -\frac{1}{g_F}) .
\end{eqnarray}
The dispersion relation of the mixture
can be found by requiring  determinant of (37) equals zero. 
\section{Results and Discussions}
In the limit of long-wavelength linear dispersion, 
the real part of the dispersion relation in absolute 
zero can be derived analytically,
\begin{eqnarray}
\omega^2 &=& 
 \left\{ \frac{1}{2}
\left(
\frac{g_B n_B}{m_B} +\frac{v_F^2}{3} 
\right)   \right. \nonumber  \\  
&& \left. \pm \sqrt{ \frac{1}{4}
\left(\frac{g_B n_B}{m_B} - \frac{v_F^2}{3} 
\right)^2 + 
\frac{ g_A^2 N(0) v_F^2n_B }{6 m_B} 
}
 \right\} k^2 ,
\end{eqnarray}
plus and minus sign in front of 
the square root corresponds 
to the two eigen modes of the mixture,
Anderson mode and Bogoliubov mode.
We can see from the last term inside the square root
that there is a repulsion
depending on average interaction between 
boson and fermion $(g_A)$. 
If the Anderson velocity exceeds Bogoliubov 
velocity, this effect stiffens the Anderson mode 
and soften the Bogoliubov mode, and vice versa.
A simple plot of  $g_A$ dependence 
to this effect is plotted 
for a superfluid mixture of fermionic 
$^6{\mathrm Li}$ (Fig.1)  
 and bosonic $^{87}{\mathrm Rb}$ (Fig.2)
as an example.   

 As the interaction becomes 
strong  enough to satisfy the condition  
\begin{eqnarray}
a_A^2 > \frac{4 \pi a_B m_R^2}{k_F m_B m_F} , 
\end{eqnarray}
the frequency of slower mode becomes a 
purely imaginary number. This corresponds
to a instability, however, phase separation
of boson and fermion occurs before this
can be observed \cite{StoofBF}.
\begin{figure}[h]
\begin{center}
\includegraphics[keepaspectratio=true, height=.4\textwidth, angle=270]%
{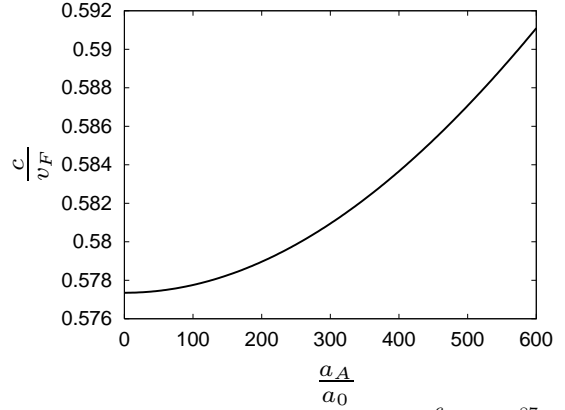}
\caption{Modification of Anderson mode in
$^6{\mathrm Li}$ and $^{87}{\mathrm Rb}$ superfluid mixture varying the
 boson-fermion interaction.
Densities of fermion and boson are taken to be $4 \times 10^{12} {\mathrm cm^{-3}}$ and
$10^{15} {\mathrm cm^{-3}} $ respectively. $a_0$ is the Bohr radius.}
\label{an}
\end{center}
\end{figure}
\begin{figure}[h]
\begin{center}
\includegraphics[keepaspectratio=true, height=.4\textwidth, angle=270]%
{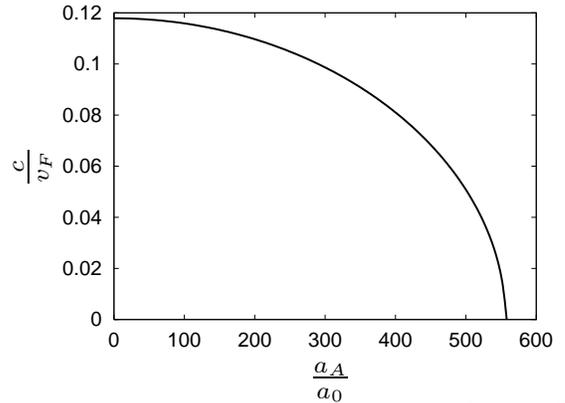}
\caption{Modification of Bogoliubov mode in
$^6{\mathrm Li}$ and $^{87}{\mathrm Rb}$ superfluid mixture varying the
 boson-fermion interaction.
Densities of fermion and boson are taken to be $4 \times 10^{12} {\mathrm cm^{-3}}$ and
$10^{15} {\mathrm cm^{-3}} $ respectively. $a_0$ is the Bohr radius.}
\label{fig:bg}
\end{center}
\end{figure}

Including the imaginary parts of bubbles,
damping of the two modes due to boson-fermion
coupling can be obtained by finding 
the numerical poles of  (37),
roots are found  to be in the form of 
$\omega = (c -i \gamma)k $  where the imaginary 
part is $k$ dependent. 
A plot of $ \gamma $ in mixture of 
$^6{\mathrm Li}$ and $^{87}{\mathrm Rb}$ is shown in Fig.3.
Note that there is no damping at absolute zero,
the reason is that  quasi-particle does not exist and therefore no
real process of quasi-particle excitation 
can take place.
The imaginary part is rather small even 
at relatively high temperature which implies that 
the two superfluid modes are stable.
However, the effect of pair breaking and Landau damping \cite{StoofBF} 
will be important near the superfluid transition temperature 
of fermion $T_{cF}$, which is out of the scope of our calculation.
 Bose-Fermi superfluid mixture near 
$T_{cF}$ will be discussed elsewhere.         
\begin{figure}[h]
\begin{center}
\includegraphics[keepaspectratio=true, height=.4\textwidth, angle=270]%
{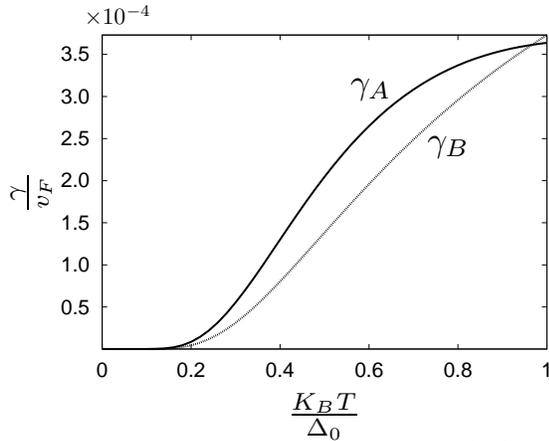}
\caption{Imaginary part of Anderson velocity $(\gamma_A)$ and Bogoliubov
 velocity  $(\gamma_B)$ for
{ $a_\uparrow = a_\downarrow = 150 a_0$} in a 
$^6{\mathrm Li}$ - $^{87}{\mathrm Rb}$ superfluid mixture with density of 
$4 \times 10^{12} {\mathrm cm^{-3}}$ and $10^{15} {\mathrm cm^{-3}} $
 respectively.}
\label{fig:im}
\end{center}
\end{figure}
\section{Conclusions}
We have calculated the dispersion relation and damping 
of a dilute mixture of boson and two component fermion.
Repulsion between the Anderson mode and Bogoliubov mode
due to boson-fermion interaction is found.
Instability of boson superfluidity is predicted 
in strong coupling regime,
where phase separation is also predicted \cite{StoofBF}.
Damping is found to be small in low temperature 
region where pair breaking effects are negligible.
\acknowledgments
This work was partly supported by COE Program ``Molecular Nano-Engineering''
from the Ministry of Education, Science and Culture, Japan.
 
\end{document}